\documentstyle[twocolumn,aps]{revtex}
\newcommand{\req}[1]{(\ref{#1})}
\newcommand{\be}{\begin{equation}}
\newcommand{\ee}{\end{equation}}
\newcommand{\bea}{\begin{eqnarray}}
\newcommand{\eea}{\end{eqnarray}}
\newcommand{\bra}{\langle}
\newcommand{\ket}{\rangle}
\newcommand{\e}{\mbox{{\rm e}}}                     
\newcommand{\Eq}{Eq.}

\newcommand{\sign}{\mbox{sgn}}
\begin{document} 
\title{\huge\bf Supersymmetric quantum mechanics \\
    with nonlocal potentials}
\author{Je-Young Choi\footnote{jychoi@kachi.yit.ac.kr }}
\address{Department of Basic Sciences \\
          Youngdong University \\
        Youngdong, Chungbuk 370-800, KOREA}
\author{Seok-In Hong\footnote{sihong@gyeyang.inchon-e.ac.kr}}
\address{Department of Science Education   \\
               Inchon National University of Education \\
               Inchon 407-753, KOREA}
\date{\today}
\maketitle
\begin{abstract}
We consider supersymmetric quantum mechanical models 
with both local and nonlocal potentials. 
We present a nonlocal deformation 
of exactly solvable local models.
Its energy eigenfunctions and eigenvalues are determined exactly.
We observe that {\em both} our model Hamiltonian and its supersymmetric partner may have 
{\em normalizable} zero-energy ground states,                       
in contrast to local models with nonperiodic or periodic potentials. 
\end{abstract}
\narrowtext
%
%
%
\section{Introduction}
  Nonlocal potentials occur in the single particle descriptions
  of many-body systems such as the Hartree-Fock approximation \cite{FW71}
  and have intrinsic applications in 
  atomic and molecular physics \cite{BJ86}, 
  solid-state physics \cite{Zi64}, and nuclear physics \cite{BBC98}.
  Nonlocal models are also used in studying statistical properties 
  of extended objects such as polymers and proteins \cite{Ni90}.
 
  In this paper we are interested in  exactly solvable quantum mechanical models 
  with nonlocal potentials.
  We resort here to supersymmetric (SUSY) quantum mechanics \cite{Wi81,CKS95} 
  which has been extensively studied recently. 
  SUSY has been applied 
  to exact solvability \cite{CGK87} in connection with shape invariance \cite{Ge83},
  inverse scattering \cite{KR86}, supersymmetry-inspired WKB method \cite{DKS86}, and 
  surface critical phenomena of inhomogeneous Ising models \cite{Ch93}.
  It is also generalized to higher dimensions \cite{DP97}, $N$-body problems \cite{GKS98},
  and dynamics \cite{BS97}.

  In SUSY quantum mechanics, a given Hamiltonian 
  and its SUSY partner have identical spectra 
  except the zero-energy ground state. 
  In ordinary cases, either one or neither of the partner Hamiltonians
  has a zero-energy ground state, in which case SUSY is unbroken or broken, respectively.
  Dunne et al.\ \cite{Du98} considered SUSY quantum mechanical models 
  with periodic potentials to find that it is possible  
  for both SUSY partners to have zero-energy ground states.
  However their ground states are {\em not} normalizable because they are Bloch states.
  In the present paper we consider SUSY quantum mechanical models 
  with both local and nonlocal potentials.
  We present a simple nonlocal deformation 
  of exactly solvable local models.
  We find the energy eigenfunctions and eigenvalues exactly and
  observe that {\em both} supersymmetric partners may have, in some parameter range,
  {\em normalizable} zero-energy ground states,
  which are in contrast to local models with nonperiodic or periodic potentials. 

\section{Nonlocal potentials and supersymmetry}  
  The time-independent Schr\"{o}dinger equation with both local 
  and nonlocal potentials in one dimension is, in the position representation,
  \bea
    \lefteqn{-\frac{\hbar^2}{2m}\frac{d^2}{dx^2}\psi(x)+V_{-}(x)\psi(x)}
                    \nonumber\\
    && \hspace{1.5cm}\mbox{}+\int_{-\infty}^{+\infty}dy\,v_{-}(x,y)\psi(y)
      =E_{-}\psi(x)\,. 
  \label{NLSch}
  \eea
  The invariance under time reversal and the conservation of the probability
  require that $V_-(x)$ and $v_-(x,y)$ are real and
  that $v_-(x,y)$ is symmetric in $x$ and $y$. 
  In the Hartree-Fock approximation the nonlocal potential (or the proper self-energy)
  is determined 
  self-consistently from the given external local potential 
  and two-body interactions \cite{FW71}.
  But in this paper, we assume that the potentials are given in the beginning.

  There are trivial choices of potentials which are exactly solvable.
  Suppose that $\psi_n(x)$ are the real normalized eigenstates 
            of a local-potential problem 
  \be
    -\frac{\hbar^2}{2m}\frac{d^2}{dx^2}\psi_n(x)+V_{-}(x)\psi_n(x)
                 =E^{(0)}_n\psi_n(x) \label{localH}
  \ee
  with $n$ the number of nodes of $\psi_n(x)$. Then 
  a trivial choice of the nonlocal potential $v_-(x,y)$ is of the form
  \be
  v_-(x,y)=
           \sum_{n}\epsilon_n \psi_n(x)\psi_n(y) \nonumber \\
  \ee
  with $\epsilon_n$ real numbers.
  Then $\psi_n(x)$ is the eigenfunction of \Eq~\req{NLSch} 
  with energy eigenvalue $E_n=E^{(0)}_n+\epsilon_n $.
  Hence we see that even though $E^{(0)}_n<E^{(0)}_{n+1}$ in local models, 
  $E_n<E_{n+1}$  may not hold in models with nonlocal potentials. 
  We may also consider a momentum dependent potential $v(\hat p)$
  \cite{hideriv}. 
  In the position representation,
  it may be alternatively represented as a nonlocal potential
  \be
    v_-(x,y)=\frac{1}{2\pi\hbar}\int_{-\infty}^{\infty}dp
                 \,\e^{i p(x-y)/\hbar}v(p) \,,
  \ee 
  which depends only on the difference $x-y$.
  If we further restrict the local potential $V_-(x)$ to be harmonic, 
  then in the momentum representation, we have
  \be
     -\frac{1}{2}m\omega^2\hbar^2\frac{d^2}{dp^2}\psi(p)
     +\left(\frac{p^2}{2m}+v(p)\right)\psi(p)=E\psi(p)\,,
  \ee
  which is of the same form as the local-potential problem \req{localH}
  in the position representation,
  so that we may exploit the solvable models of local potentials.
  In addition, nonlocal separable potentials also lead to solvable models \cite{Gla77}. 
  In the following we will exclude these cases.

  Sometimes it is convenient to write potentials in operator form as
  \bea
    \hat{V}_-&=&\int_{-\infty}^{\infty}dx\,|x\ket V_-(x)\bra x|\,, \nonumber\\
    \hat{v}_-&=&\int_{-\infty}^{\infty}dx\!\!\int_{-\infty}^{\infty}dy\,
                      |x\ket v_-(x,y)\bra y|  \,.
  \label{potop}
  \eea
  Then the Hamiltonian is
  \be
     \hat H_-=\frac{\hat{p}^2}{2m}+\hat{V}_-+\hat{v}_-\,.
  \ee
  As in SUSY quantum mechanics with local potentials \cite{CKS95},
  we assume that the Hamiltonian $\hat H_-$ 
  can be factorized in the form $\hat{H}_{-}=\hat{A}^{\dagger}\hat{A}$.
  Its SUSY partner Hamiltonian $\hat H_+$ is then given by
  \be
    \hat{H}_{+}=\hat{A}\hat{A}^{\dagger}
              =\frac{\hat{p}^2}{{2m}}+{\hat V}_+ +{\hat v}_+ \,.
  \ee
  First-order differential operators $\hat{A}^{\dagger}$ and $\hat{A}$ are defined by
  \bea
    \hat{A}^{\dagger}&=&-\frac{i \hat{p}}{\sqrt{2m}}+\hat{W}+\hat{w} \,,
                          \nonumber\\
    \hat{A}&=&\frac{i \hat{p}}{\sqrt{2m}}+\hat{W}+\hat{w} \,,
    \label{AAdagger}   
  \eea
  where $\hat{W}$ and $\hat{w}$ are defined by \Eq~\req{potop}
  with $V_-$ and $v_-$ replaced by the local superpotential $W$ 
  and the nonlocal superpotential $w$, respectively.
  Then we see that the potentials are written in terms of the superpotentials 
   $W(x)$ and $w(x,y)$ as \
  \bea
    V_{\pm}(x)&=&[W(x)]^2\pm\frac{\hbar}{\sqrt{2m}}\frac{dW(x)}{dx}\,,
                 \nonumber  \\
    v_{\pm}(x,y)&=&\int_{-\infty}^{\infty}du \,w(x,u)w(u,y)
                      +\{W(x)+W(y)\}w(x,y)\nonumber  \\
          & &  {}\pm\frac{\hbar}{\sqrt{2m}}
                     \left[\frac{\partial w(x,y)}{\partial x}
                      +\frac{\partial w(x,y)}{\partial y} \right]\,.  
    \label{wtov}
  \eea
  If $w(x,y)$ is symmetric, then so are $v_{\pm}(x,y)$.
  If $W(x)$ and $w(x,y)$ are real, so are $V_{\pm}(x)$ and $v_{\pm}(x,y)$.
 
  Whereas in local models the relationship between the ground state 
  and the superpotential $W(x)$ is one-to-one, 
  this is not true in nonlocal models.
  Given the superpotentials $W(x)$ and $w(x,y)$ 
  the zero-energy ground state $\psi_0(x)$ 
  of $\hat H_-$ is obtained from the integro-differential equation
  \be
     \frac{\hbar}{\sqrt{2m}}\frac{d\psi_0(x)}{dx}+W(x)\psi_0(x)
        +\int_{-\infty}^{\infty}dy \,w(x,y)\psi_0(y)=0  \,.
  \ee 
  Conversely, for any normalized function $\psi_0(x)$, 
  we could choose 
  \bea
  \lefteqn{w(x,y)=-\frac{\hbar}{\sqrt{2m}}
           \left[\psi'_0(x)\psi_0(y)+\psi_0(x)\psi'_0(y)\right]}\nonumber\\
    &&  \hspace{2cm} \mbox{}-\psi_0(x)\{W(x)+W(y)\}\psi_0(y)   \nonumber\\
    &&  \hspace{2cm} \mbox{}+C(x,y)\,,
  \eea  
  Then $\psi_0(x)$ is the zero-energy ground state as long as 
  arbitrary real functions $W(x)$ and $C(x,y)$ [$=C(y,x)$] satisfy the requirements
  \bea
    \int_{-\infty}^{\infty}dy \,W(y)\,[\psi_0(y)]^2&=&0 \,,  \nonumber\\
    \int_{-\infty}^{\infty}dy \,C(x,y)\,\psi_0(y) &=&0 \,.
  \eea
  In order to determine higher energy eigenvalues exactly,
  we need to specify the superpotentials.

\section{A nonlocal deformation of local models}
  As a specific example of the above formalism, we construct 
  a class of exactly solvable models
  with both local and nonlocal potentials
  starting from any exactly solvable local model with a superpotential $W_{0}(x)$
  which is an odd function of $x$\,.
  Examples are the harmonic oscillator with $W_0(x)=\sqrt{\frac{m}{2}}\omega x$
  and the Rosen-Morse II or Scarf II potential with $W_0(x)=a \tanh\alpha x$. 
  Now we choose
  \bea
    W(x)&=&(1-c)W_{0}(x)\,,\nonumber\\
    w(x,y)&=&-c{{\hbar}\over\sqrt{2m}}\delta'(x+y)\,, 
  \eea
  where $c$ is a real parameter of nonlocality.  
  Using \Eq~\req{wtov} we obtain that
  \bea
    \lefteqn{V_{\pm}(x)=(1-c)^2[W_0(x)]^2\pm (1-c)\frac{\hbar}{\sqrt{2m}}W'_0(x)\,,}
                                             \nonumber\\
    \lefteqn{\int_{-\infty}^{\infty} dy\, v_{\pm}(x,y)\psi_{\pm}(y)}\nonumber\\
         &&\hspace{1cm}=\mbox{}c^2{\hbar^2\over 2m}\psi''_{\pm}(x)
                       \mp2c{\hbar^2\over2m}\psi''_{\pm}(-x)  \nonumber\\
        &&\hspace{1.5cm}\mbox{}
          +c\left(1-{c}\right){{\hbar}\over\sqrt{2m}}W'_0(x)\psi_{\pm}(-x)\,.
  \eea
  The nonlocal potential may be written as
  \be 
    {\hat v}_{\pm}=c^2{{\hat p}^2\over 2m}\pm 2c{{\hat p}^2\over2m}\hat P
           +c\left(1-{c}\right){{\hbar}\over\sqrt{2m}}W'_0({\hat x})\hat P \,,
  \ee
  where $\hat P$ is the parity operator.
  Thus the Hamiltonians $\hat H_s$ with $s=\pm$ are written as 
  \bea
    \lefteqn{\hat H_s= (1+s\hat P c)^2\frac{\hat p^2}{2m}  
        +(1-c)^2[W_0(\hat x)]^2}\nonumber\\
    &&\hspace{1.4cm}\mbox{}+s(1-c)\frac{\hbar}{\sqrt{2m}}W'_0(\hat x)(1+s\hat P c)\,.
  \eea
  We may mention that there is no classical analog of the present model
  since $\hat H_{\pm}$ contains the parity operator. 

  Since $[\hat H_{\pm},\hat P]=0$,
  we look for simultaneous eigenstates $\psi_{\pm}(x)$ of $\hat H_{\pm}$ and $\hat P$. 
  If ${\hat P}\psi_{s}(x)=\psi_{s}(-x)=P\psi_{s}(x)$ with $P=\pm1$,
  then the eigenvalue equation for $\hat H_{s}$ becomes
  \widetext
  \be
    -(1+sPc)^2\frac{\hbar^2}{2m}\psi_{s}''(x)  
        +\left[(1-c)^2[W_0(x)]^2+s(1+sPc)(1-c)\frac{\hbar}{\sqrt{2m}}W'_0(x)
                \right]\psi_{s}(x)=E_{s}\psi_{s}(x) \,.  
  \ee
  \narrowtext

  There are two singular cases ($c=\pm1$). 
  At $c=1\,$, $\hat H_{\pm}$ 
  in the $P={\pm}1$ sector
  is a free Hamiltonian,
 which has no bound state.
  At $c=\pm1\,$, $\hat H_-$ ($\hat H_+$) has no kinetic term
  in the $P=\pm1$ ($P=\mp1$) sector.
  In the following we will disregard these cases.

  We may represent ${\hat H}_s$ 
  in terms of the local Hamiltonian ${\hat H}_s^{(0)}$ 
  ($={\hat H}_s$ with $c=0$) as follows:\cite{footnote1}
  \be
  \hat H_s=\left\{\begin{array}{ll}
             (1-c)^2\hat H_s^{(0)} &\hspace{1cm}
(P=-s) \\
               \left.(1-c)^2\hat H_{\alpha s}^{(0)}
              \right|_{\hbar\rightarrow\hbar(c)} &\hspace{1cm}
(P=+s)\,,
       \end{array} \right.
    \label{HHzero}
  \ee
  where the symbols $\alpha=\sign[(1+c)/(1-c)]$ and 
  $\hbar(c)=\hbar|(1+c)/(1-c)|$ are introduced.  
  We have $\alpha=+1$ for $|c|<1$ and $\alpha=-1$ for $|c|>1$.
  Note that the local Hamiltonians ${\hat H}_s^{(0)}$ are even in parity.
  The above relation \req{HHzero} may be summarized in Table \ref{tbl}.
  From the table or \Eq~\req{HHzero}, we see, in particular, 
  that for any $|c|>1$ ($\alpha=-1$), both $\hat H_+$ and $\hat H_-$
  in the given parity sector $P$ are proportional to $\hat H_{-P}^{(0)}$
  (up to an appropriate rescaling of $\hbar$, when necessary),
  and thus both operators have normalizable zero-energy ground states. 

  We assume  $\sign W_0(\pm\infty)=\pm1$, 
  which ensures the existence of unbroken SUSY 
  for local models ($c=0$) where $\hat H_-^{(0)}$ has a zero-energy ground state.
  Then the normalized eigenstates $\psi_{s,n}^{(0)}(x)$ of $\hat H_s^{(0)}$ 
  have energies $E_{s,n}^{(0)}$, 
  where the nonnegative integer $n$ labels the number of nodes: 
  $E_{-,0}^{(0)}=0$, $E_{-,n}^{(0)}=E_{+,n-1}^{(0)}\,$, and $P=(-1)^n$.
  The eigenstates of $\hat H_-^{(0)}$ and $\hat H_+^{(0)}$are related by
  \bea
    {\hat A}^{(0)}\psi_{-,n}^{(0)}&=& 
                \sqrt{E_{-,n}^{(0)}}\,\psi_{+,n-1}^{(0)}\,,\nonumber\\ 
    {\hat A}^{(0)\dagger}\psi_{+,n}^{(0)}&=&
                \sqrt{E_{+,n}^{(0)}}\,\psi_{-,n+1}^{(0)}\,,
  \eea
  where ${\hat A}^{(0)}$ and ${\hat A}^{(0)\dagger}$ are given by \Eq~\req{AAdagger}
  with $\hat W=\hat W_0$ and $\hat w=0$.
  Now we see that the normalized eigenfunctions of $\hat H_{\pm}$ are given by
  \bea
     \psi_{-,n}(x)&=&\left\{\begin{array}{ll}\psi_{-,n}^{(0)}(x)  
                      &\hspace{.3cm}\mbox{($n$ even, $c\neq1$)}  \\
              \left.\psi_{-,n}^{(0)}(x)
                    \right|_{\hbar\rightarrow\hbar(c)}  
                      &\hspace{.3cm}\mbox{($n$ odd, $|c|<1$)}  \\
             \left.\psi_{+,n}^{(0)}(x)
                    \right|_{\hbar\rightarrow\hbar(c)}  
                      &\hspace{.3cm}\mbox{($n$ odd, $|c|>1$)}\,,  
              \end{array}\right.             \nonumber\\
     \psi_{+,n}(x)&=&\left\{\begin{array}{ll}\psi_{+,n}^{(0)}(x)  
                      &\hspace{.3cm}\mbox{($n$ odd, $c\neq1$)}  \\
              \left.\psi_{+,n}^{(0)}(x)
                    \right|_{\hbar\rightarrow\hbar(c)}  
                      &\hspace{.3cm}\mbox{($n$ even, $|c|<1$)}  \\
             \left.\psi_{-,n}^{(0)}(x)
                    \right|_{\hbar\rightarrow\hbar(c)}  
                      &\hspace{.32cm}\mbox{($n$ even, $|c|>1$)}\,.  
              \end{array}\right.  
  \eea  
  The corresponding energy eigenvalues are given by
  \bea
    E_{-,n}&=&\left\{\begin{array}{ll}
                (1-c)^2E_{-,n}^{(0)} 
                         &\hspace{.3cm}\mbox{($n$ even, $c\neq1$)}  \\
       \left.(1-c)^2 E_{-,n}^{(0)}\right|_{\hbar\rightarrow\hbar(c)}  
                               &\hspace{.3cm}\mbox{($n$ odd, $|c|<1$)}\\
       \left.(1-c)^2 E_{+,n}^{(0)}\right|_{\hbar\rightarrow\hbar(c)} 
                               &\hspace{.3cm}\mbox{($n$ odd, $|c|>1$)}\,,
              \end{array}\right.         \nonumber\\
    E_{+,n}&=&\left\{\begin{array}{ll}
                (1-c)^2E_{+,n}^{(0)}    
                            &\hspace{.3cm}\mbox{($n$ odd, $c\neq1$)}  \\
       \left.(1-c)^2 E_{+,n}^{(0)}\right|_{\hbar\rightarrow\hbar(c)} 
                              &\hspace{.3cm}\mbox{($n$ even, $|c|<1$)}\\
       \left.(1-c)^2E_{-,n}^{(0)}\right|_{\hbar\rightarrow\hbar(c)}  
                            &\hspace{.3cm}\mbox{($n$ even, $|c|>1$)}\,.
              \end{array}\right.  
  \eea

  As in local models we have labelled the eigenfunctions by the number of their nodes.
  But in the nonlocal models, the energy eigenvalues are not ordered 
  by the number of nodes. 
  The operator $\hat A$ (${\hat A}^{\dagger}$) converts an eigenfunction 
  of $\hat H_-$ ($\hat H_+$) into an eigenfunction of $\hat H_+$ ($\hat H_-$) 
  with the same energy. 
  Explicitly,
  \bea
     \hat A \psi_{-,n}(x)&=&\left\{\begin{array}{ll}
         A_{-,n}\,\psi_{+,n-1}(x)
                 & \hspace{.3cm}\mbox{($n$ even, $c\neq1$)}         \\
           A_{-,n}\,\psi_{+,n-1}(x)
                 &\hspace{.3cm}\mbox{($n$ odd, $|c|<1$)}           \\
         A_{-,n}\,\psi_{+,n+1}(x)
                 &\hspace{.3cm}\mbox{($n$ odd, $|c|>1$)} \,,         
              \end{array}\right.      \nonumber\\
     {\hat A}^{\dagger} \psi_{+,n}(x)&=&\left\{\begin{array}{ll}
         A_{+,n}\,\psi_{-,n+1}(x)
                 & \hspace{.3cm}\mbox{($n$ odd, $c\neq1$)}        \\
           A_{+,n}\,\psi_{-,n+1}(x)
                 & \hspace{.3cm}\mbox{($n$ even, $|c|<1$)}        \\
         A_{+,n}\,\psi_{-,n-1}(x)
                 &\hspace{.3cm}\mbox{($n$ even, $|c|>1$)}\,. 
              \end{array}\right.
  \eea 
  where $A_{s,n}=\sign{(1-c)}\sqrt{E_{s,n}}\,$.
  We see that $\hat A\,$, when applied to eigenfunctions of $\hat H_-$,
  destroys a node in the even eigenfunction 
  for $c\neq 1$ or in the odd eigenfunction for $|c|<1$, 
  but creates an extra node in the odd eigenfuction for $|c|>1$.
  Similar behavior holds for the operator $\hat A^{\dagger}$.
  If $|c|<1$, then the ground state $\psi_{+,0}(x)$ of $\hat H_+$ 
  has a positive energy and the SUSY is unbroken 
  (Witten index $\Delta=1-0=1$) \cite{Wi81}.
  Howerver, if $|c|>1$, then the ground states $\psi_{-,0}(x)$ and 
  $\psi_{+,0}(x)$ both have zero energy 
  as discussed above,
  and are annihilated by ${\hat A}$ and by ${\hat A}^{\dagger}$, respectively. 
  The SUSY is still unbroken for $|c|>1$ 
  even though the Witten index $\Delta=1-1=0$. 
  This case is similar to that with periodic potentials considered by
  Dunne et al.\ \cite{Du98} who found that it is possible  
  for both SUSY partners to have zero-energy ground states.
  Their ground states are {\em not} normalizable because they are Bloch states.
  However, normalizable are the eigenfunctions we obtained in the present paper.
  
  It may be noted that at particular values of $c$ some of energy eigenvalues are doubly degenerate. 
  In these cases the Hamiltonian $\hat H_{-}$ (or $\hat H_{+}$) 
  and the parity operator $\hat P$ form 
  a complete set of commuting observables 
  in contrast to the normal situation in one dimension,
  where the Hamiltonian alone is sufficient.

\section{Conclusions}
  We have considered nonlocal potentials 
  in connection with SUSY quantum mechanics.
  The local and nonlocal potentials are expressed in terms of
  the local and nonlocal superpotentials.
  We have presented a nonlocal extension 
  of exactly solvable local models.
  Its energy eigenfunctions and eigenvalues are determined exactly
  and it is observed that {\em both} supersymmetric partners ($\hat H_{\pm}$)
  may have, in some parameter range,
  {\em normalizable} zero-energy ground states.
  It would be interesting to find a (sufficient) condition of solvability 
  generalizing the shape invariance 
  which is applicable to models with both local and nonlocal potentials.

\begin{table}
\caption{The representation of the Hamiltonians $\hat H_s$ in the parity sector $P=\pm1$
     in terms of the local Hamiltonians $\hat H_s^{(0)}$ ($=\hat H_s$ with $c=0$).
     The symbols $\alpha$ and $\hbar(c)$ are defined in the text.}
\begin{tabular}{c|c|c}
      & $P=+1$ & $P=-1$ \\ \hline 
  $\hat H_-$ & $(1-c)^2\hat H_-^{(0)}$ 
       & $\left.(1-c)^2\hat H_{-\alpha}^{(0)}\right|_{\hbar\rightarrow\hbar(c)}$\\ \hline
  $\hat H_+$  
       & $\left.(1-c)^2\hat H_{\alpha}^{(0)}\right|_{\hbar\rightarrow\hbar(c)}$
       & $(1-c)^2\hat H_+^{(0)}$
\end{tabular}
\label{tbl}
\end{table}

\begin{thebibliography}{99}
\bibitem{hideriv} In the position representation 
    it may also be thought of as higher derivative terms 
   $v(\frac{\hbar}{i}\frac{d}{dx})$. 
     The invariance under time reversal and the conservation of the probability
  require that $v(\hat p)$ must be Hermitian and even in $\hat p$.
\bibitem{footnote1}
              Here it is assumed that $W_0(x)$ is independent of $\hbar$.
\end{thebibliography}
\end{document}